\documentstyle[12pt,twoside,fleqn,espcrc1,epsf]{article}

\newcommand{\AmS}{{\protect\the\textfont2
  A\kern-.1667em\lower.5ex\hbox{M}\kern-.125emS}}

\title{Deeply virtual electroproduction of photons and mesons 
  on the nucleon}

\author{M. Vanderhaeghen\address{Institut f\"ur Kernphysik, University
        of Mainz, D-55099 Mainz, Germany }
        , P.A.M. Guichon\address{Service de Physique Nucl\'eaire,
        CEA/Saclay, F-91191 Gif-sur-Yvette, France}
        and M. Guidal\address{Institut de Physique Nucl\'eaire - IN2P3,
        F-91406 Orsay, France}
       }

\begin{document}
\maketitle

\begin{abstract}
We give predictions for the leading order amplitudes for deeply virtual
Compton scattering and hard meson electroproduction reactions at large
$Q^2$ in the valence region in terms of skewed quark distributions. 
We give first estimates for the power corrections to these
leading order amplitudes. In particular, we outline examples of 
experimental opportunities to access the skewed parton distributions 
at the current high-energy lepton facilities~: JLab, HERMES and COMPASS.
\end{abstract}

\section*{}


In recent years,   
a unified theoretical description of a wide variety of exclusive
processes in the Bjorken regime has emerged through the
formalism introducing new generalized parton distributions, the so-called
skewed parton distributions (SPD's). It has been
shown that these distributions, which parametrize the structure 
of the nucleon, allow to describe in leading order perturbative
QCD (PQCD), various exclusive processes such as, in particular,
deeply virtual Compton scattering (DVCS) and longitudinal 
electroproduction of vector and pseudoscalar mesons
(see e.g. Refs.\cite{Ji97}-\cite{vcsrev} and references therein).

\begin{figure}[ht]
\vspace{-.75cm}
\epsfxsize=11 cm
\epsfysize=7. cm
\centerline{\epsffile{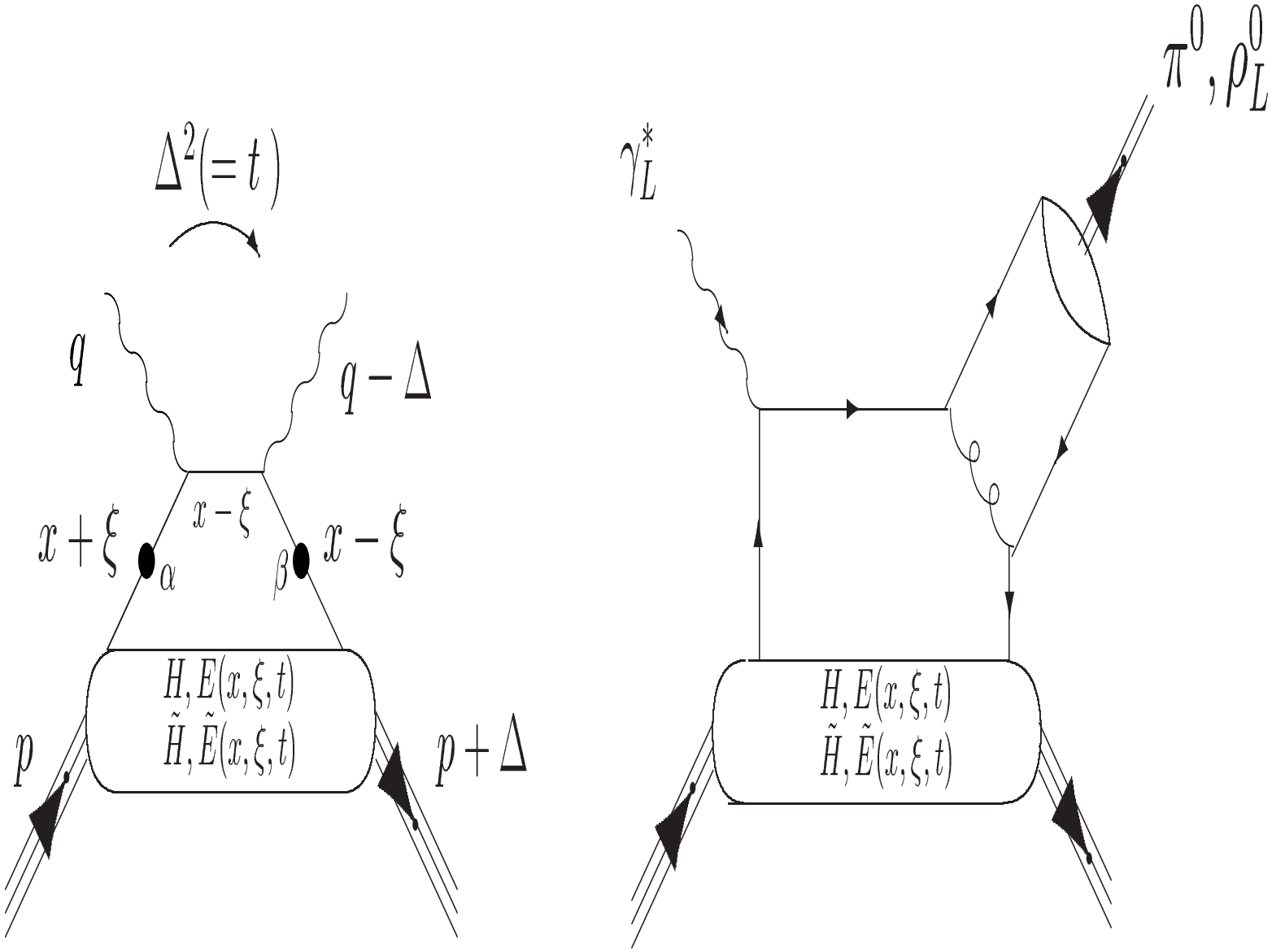}}
\vspace{-3.25cm}
\caption[]{Leading order diagrams for DVCS (left) and for longitudinal
  electroproduction of mesons (right).}
\label{fig:handbags}
\vspace{-1cm}
\end{figure}

The leading order PQCD diagrams for DVCS
and hard meson electroproduction are of the type as shown in 
Fig.~\ref{fig:handbags}.
It has been proven \cite{Ji97,Rady96} 
that the leading order DVCS amplitude in
the forward direction can be factorized in a hard scattering part 
(which is exactly calculable in PQCD) and a soft, nonperturbative nucleon 
structure part as is illustrated on the left panel of 
Fig.~\ref{fig:handbags}. 
The nucleon structure information  
can be parametrized, at leading order, in 
terms of 4 generalized structure functions. In the notation of Ji,
these functions are the off-forward parton distributions 
(OFPD's) denoted as $H, \tilde H, E, \tilde E$
which depend upon three variables : $x$, $\xi$ and $t$.
The light-cone momentum fraction $x$ is defined by $k^+ = x P^+$,
where $k$ is the quark loop momentum and  
$P$ is the average nucleon momentum (using the definition 
$a^{\pm} \equiv 1/\sqrt{2} (a^0 \pm a^3)$). 
The skewedness variable $\xi$ is 
defined by $\Delta^+ = - 2 \xi \, P^+$, where $\Delta$ is the 
overall momentum transfer in the process and where 
$2 \xi \rightarrow x_B/(1 - x_B/2)$ in the Bjorken limit. 
Furthermore, the third variable entering the OFPD's 
is given by the Mandelstam invariant $t = \Delta^2$.  
In Fig.~\ref{fig:handbags}, the variable $x$ runs from -1 to 1.
As noted by Radyushkin \cite{Rady96}, one can identify two regions for
the SPD's. In the regions where $x > \xi$ or  $x < - \xi$, 
the SPD's are the generalizations of the usual parton distributions from 
DIS. Actually, in the forward direction, the OFPD's $H$ and $\tilde H$ 
respectively reduce to the quark density distribution $q(x)$ and 
quark helicity distribution $\Delta q(x)$, obtained from DIS. 
In the region $ -\xi < x < \xi$, the SPD's 
behave as a ``meson-like'' distribution amplitude and contain new
information about nucleon structure \cite{Poly99}.  

To provide estimates for electroproduction observables, we need a
model for the SPD's. 
The following calculations were performed using $\xi$-dependent
SPD's based on a product ansatz (for the double distributions 
\cite{Rady98}) of a quark distribution (we use the MRST98 quark distributions
\cite{MRST98} as input) 
and an asymptotic ``meson-like'' distribution amplitude 
(see Ref.~\cite{VGG99} for more details). The $t$-dependence is given
by the corresponding form factors (Dirac form factor for $H$, 
axial form factor for $\tilde H$). 

Besides the DVCS process, 
a factorization proof was also given for the leading
order meson electroproduction amplitude in the valence region at large
$Q^2$ \cite{Collins97}, which is shown on the right panel of 
Fig.~\ref{fig:handbags}. 
This factorization theorem only applies when the virtual photon is
\underline{longitudinally} polarized. 
The leading order longitudinal amplitude for meson electroproduction 
behaves as $1/Q$. This leads to a $1/Q^6$ scaling behavior for the
longitudinal cross section $d \sigma_L / d t$ at large $Q^2$, which
provides an experimental signature of the leading order mechanism. 
Besides the dependence on the SPD's, the meson electroproduction
amplitudes require the additional non-perturbative input from the 
meson distribution amplitudes,  for which we take
asymptotic forms in the calculations. 
Furthermore, it was shown in Ref.\cite{Collins97} that
electroproduction of vector (pseudoscalar) 
mesons accesses the unpolarized (polarized) OFPD's $H$
and $E$ ($\tilde H$ and $\tilde E$) respectively.
According to the produced meson  
($\rho^0$, $\rho^\pm$, $\omega$, $\phi$, $\pi^0$, $\pi^\pm$,
$\eta$,...), the SPD's for the different
quark flavors enter in different combinations due to the different 
quark charges and isospin factors. 


\begin{figure}[h]
\vspace{-2cm}
\hspace{.5cm}
\epsfxsize=12 cm
\epsfysize=13 cm
\centerline{\epsffile{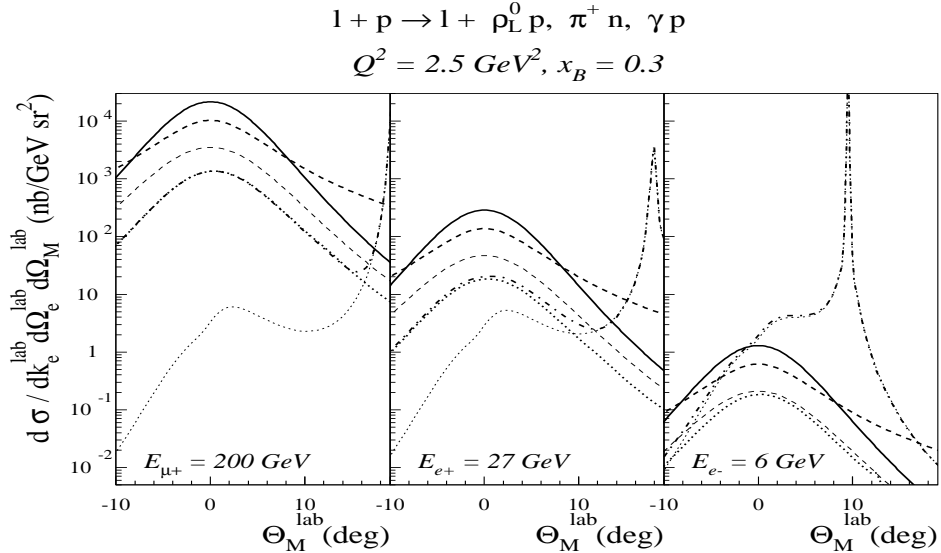}}
\vspace{-4.5cm}
\caption[]{\small Comparison between the angular dependence of the 
leading order predictions for 
the $\rho^0_L$ (full lines), $\pi^+$ (thick, upper dashed lines) 
leptoproduction and DVCS (thick dotted lines) 
in-plane cross sections at $Q^2$ = 2.5 GeV$^2$, $x_B$ = 0.3 
and for different beam energies : 
$E_{\mu^+}$ = 200 GeV (COMPASS), $E_{e^+}$ = 27 GeV (HERMES), 
$E_{e^-}$ = 6 GeV (JLab). The BH (thin dotted lines) 
and total $\gamma$ (dashed-dotted lines) cross sections are also
shown. For the $\pi^+$, the result excluding the pion pole is also
shown (thin, lower dashed lines).}
\label{fig:diffkin}
\vspace{-.5cm}
\end{figure}

We show here some results for DVCS and meson
electroproduction observables using the $\xi$-dependent 
ansatz for the OFPD's described previously. For more results, 
we refer to our works in Refs.\cite{vcsrev,VGG98,VGG99}.
In Fig.~\ref{fig:diffkin}, 
the $\rho^0_L$, $\pi^+$ and $\gamma$ cross sections 
are compared as function of 
the beam energy at fixed $Q^2$ in the valence region. 
Going up in energy, the increasing virtual photon flux factor boosts the 
meson leptoproduction cross sections and the DVCS part of 
the $\gamma$ leptoproduction cross section.
Comparing the different channels, 
it is clear on this picture that the $\rho^0_L$ channel is very
favorable as it depends on the unpolarized SPD's. 
For the $\gamma$ channel, the contaminating Bethe-Heitler (BH) process 
is hardly influenced by the beam 
energy and therefore overwhelms the DVCS cross section at low beam energies. 
Although Fig.~\ref{fig:diffkin} shows that 
a high energy such as planned at COMPASS is  preferable, one 
can try  to undertake a preliminary study of the hard electroproduction 
reactions using the existing facilities such as HERMES or JLAB, 
despite their low energy. 
Recently, an experiment has been approved at JLAB \cite{propo98107} 
to investigate (the onset of) the scaling behavior for $\rho^0_L$
electroproduction in the valence region 
($Q^2 \approx 3.5$ GeV$^2$, $x_B \approx 0.3$). 

\begin{figure}[h]
\epsfxsize=8 cm
\epsfysize=12. cm
\centerline{\epsffile{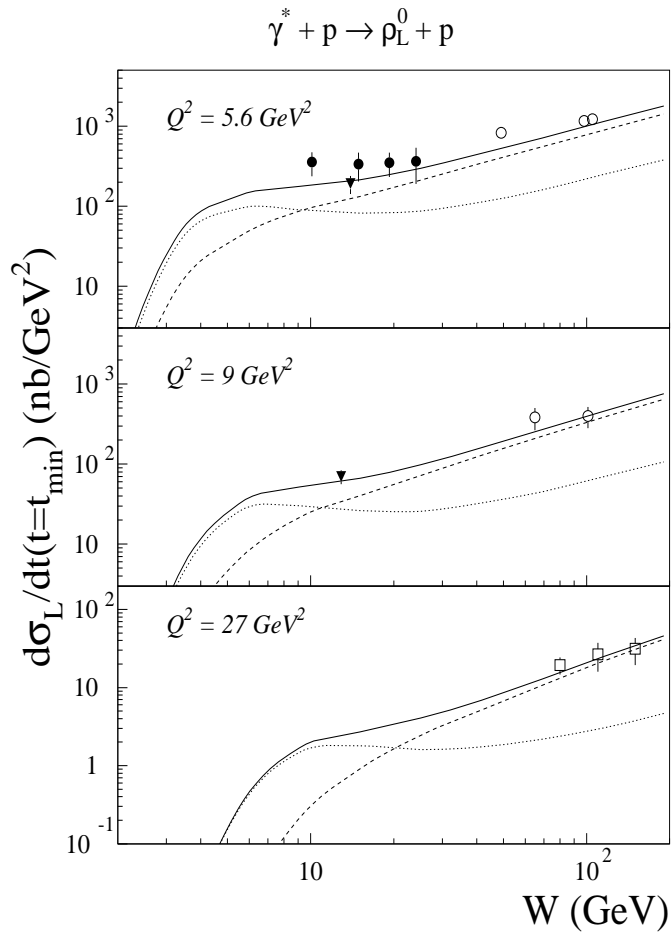}}
\vspace{-1cm}
\caption[]{\small Longitudinal forward differential cross section for 
$\rho^0_L$ electroproduction. Calculations compare the quark exchange
mechanism (dotted lines) with the two-gluon
exchange mechanism (dashed lines) and the sum of both (full
lines). Both calculations include the corrections due to intrinsic
transverse momentum dependence. 
The data are from NMC (triangles), E665 (solid circles) and 
ZEUS (open circles and squares). References 
to the data can be found in Ref.~\cite{VGG99}.}
\label{fig:rhotot2}
\end{figure}

Although at present, 
no experimental data for the \( \rho ^{0}_{L} \) electroproduction
at larger \( Q^{2} \) exist in the valence region (\( x_{B}\approx  \)
0.3), the reaction \( \gamma ^{*}\, p\longrightarrow \rho ^{0}_{L}\, p \)
has been measured at smaller values of \( x_{B} \). 
We therefore compare our results in Fig.~\ref{fig:rhotot2} 
to see how these data approach the valence region, where one is sensitive
to the quark SPD's. For the purpose of this discussion, we call the mechanism
which proceeds through the quark SPD's, the Quark Exchange Mechanism (QEM). 
Besides the QEM, $\rho^0$ electroproduction at large $Q^2$ and small
$x_B$ proceeds predominantly through a perturbative two-gluon exchange
mechanism (PTGEM) as studied in Ref.~\cite{Fra96}. To compare to the
data at intermediate $Q^2$, we implemented in both mechanisms the
power corrections due to the parton's intrinsic transverse momentum
dependence (see Ref.~\cite{VGG99} for details), which give a
significant reduction at the lower $Q^2$ values. Comparing our results
with the data in Fig.~\ref{fig:rhotot2}, one sees that the PTGEM
explains well the fast increase at high c.m. energy ($W$) of the cross section
but substantially underestimates the data at lower energies. This is
where the QEM is expected to contribute since $x_B$ is then in the
valence region. The QEM describes well the change of behavior of the
data at lower $W$. This has also been confirmed by recent HERMES data
(around $W \approx$ 5 GeV) \cite{hermes} 
to which we compared our calculations. 

It is clear that at present new and accurate data for these exclusive
channels are needed. The fundamental interest of the SPD's justifies 
an effort towards their experimental determination in order to 
open up a new domain in the study of the nucleon structure.

\end{document}